\documentclass[12pt]{iopart}
\usepackage{graphicx}
\usepackage{appendix}
\usepackage{xcolor}
\usepackage{cite}
\usepackage{fixltx2e}

\begin{document}
\title{Massively Scalable Wavelength Diverse Integrated Photonic Linear Neuron}

\author{Matthew van Niekerk$^{1,\dagger}$, Anthony Rizzo$^2$, Hector Rubio Rivera$^1$, Gerald Leake$^3$, Daniel Coleman$^3$, Christopher Tison$^4$, Michael Fanto$^4$, Keren Bergman$^2$, and Stefan Preble$^1$}

\address{$^1$ Microsystems Engineering, Rochester Institute of Technology, Rochester, NY, USA}
\address{$^2$ Department of Electrical Engineering, Columbia University, New York, NY, USA}
\address{$^3$ State University of New York Polytechnic Institute, Albany, NY, USA}
\address{$^4$ Information Directorate, Air Force Research Laboratory, Rome, NY, USA}
\ead{$^{\dagger}$mv7146@rit.edu}

\begin{abstract}
As computing resource demands continue to escalate in the face of big data, cloud-connectivity and the internet of things, it has become imperative to develop new low-power, scalable architectures. Neuromorphic photonics, or photonic neural networks, have become a feasible solution for the physical implementation of efficient algorithms directly on-chip. This application is primarily due to the linear nature of light and the scalability of silicon photonics, specifically leveraging the wide-scale complementary metal-oxide-semiconductor (CMOS) manufacturing infrastructure used to fabricate microelectronics chips. Current neuromorphic photonic implementations stem from two paradigms: wavelength coherent and incoherent. Here, we introduce a novel architecture that supports coherent \textit{and} incoherent operation to increase the capability and capacity of photonic neural networks with a dramatic reduction in footprint compared to previous demonstrations. As a proof-of-principle, we experimentally demonstrate simple addition and subtraction operations on a foundry-fabricated silicon photonic chip. Additionally, we experimentally validate an on-chip network to predict the logical 2-bit gates AND, OR, and XOR to accuracies of $96.8\%, 99\%,$ and $98.5\%$, respectively. This architecture is compatible with highly wavelength parallel sources, enabling massively scalable photonic neural networks. 
\end{abstract}


\section{Introduction}
In recent years, integrated photonics has emerged as a viable and highly appealing candidate for neural network hardware implementation. Photonic neural networks (PNN) hold particular promise compared to traditional electronic implementations due to the highly favorable physical properties of photons compared to electrons for performing linear operations; namely, (i) the bosonic nature of light permits high degrees of parallelism in spatial mode and wavelength, (ii) such linear transformations are naturally implemented using standard photonic elements (i.e., beamsplitters and phase shifters) at the speed of light, and (iii) these operations can consume orders of magnitude less energy than their electronic counterparts \cite{shen2017deep,totvic2020femtomac, shastri2021photonics, nahmias2019photonic, cheng2020siphcodesign}. In particular, silicon photonics provides an ideal platform for PNNs due to its compatibility with the standard complementary metal-oxide-semiconductor (CMOS) infrastructure used to fabricate microelectronic chips and the high index contrast between silicon and silicon dioxide which permits unparalleled device density \cite{shastri2021photonics}. Numerous approaches to optical neural networks have been demonstrated in various platforms, including as spiking \cite{nahmias2013leaky, shastri2014photonic}, phase-change materials \cite{cheng2017chip, wu2021programmable, feldmann2021parallel}, metal-sulfide microfibers \cite{gholipour2015amorphous, sun2018optoelectronic}, photoelectric multiplication \cite{hamerly2019large, feldmann2021parallel}, and semiconductor optical amplifiers \cite{shi2019deep, fouskidis2020reconfigurable}. However, two primary categories have emerged for PNNs in the silicon photonics platform: wavelength coherent and incoherent \cite{shastri2021photonics}. 

Wavelength coherent designs rely on careful placement and tuning of Mach-Zehnder interferometers (MZI), such that optical interference provides the multiplication and summation operations. Carefully designed MZI mesh circuits can implement any real-valued matrix on-chip by reconfiguring phase shifters according to singular value decomposition \cite{shen2017deep, zhang2021optical}. The coherent optical linear neuron (COLN) architecture implements vector-vector multiplication and summation by leveraging MZIs as inputs and weights for multiplication and coherent interference as the summation operator \cite{mourgias2019neuromorphic, totovic2022programmable}.
Wavelength incoherent designs rely on multi-wavelength operation, often employing a broadband photodetector to perform \textcolor{black}{summation, but through this method the phase of the signal is lost to optical-electrical conversion}. For example, in the ``Broadcast \& Weight'' protocol, multiple wavelength channels are launched as inputs along the same waveguide, which allows for parallel weighting with micro-ring resonator weight banks \cite{tait2014broadcast}. A balanced photodetector pair sums the vector-vector multiplication; however, multiple micro-ring weight banks can operate in parallel for vector-matrix operation, as demonstrated in \cite{tait2017neuromorphic}. 

In this work, we present a novel architecture that leverages both coherent \textit{and} incoherent aspects to create a massively scalable PNN specific linear operator. This architecture requires fewer components, provides orders-of-magnitude footprint reduction, and consumes substantially less power than comparable designs. As an initial proof-of-principle, we experimentally demonstrate simple addition and subtraction on-chip. Additionally, we implement and demonstrate a neural network task on-chip, designed to perform logic gate operations (AND, OR, and XOR). Due to the compact footprint, low energy consumption, and scalability of the demonstrated linear neuron, these results pave the way toward large-scale PNNs with hundreds to thousands of neurons on a single silicon chip. 

\section{Proposed Architecture}
\begin{figure}
    \centering
    \includegraphics[width = \textwidth]{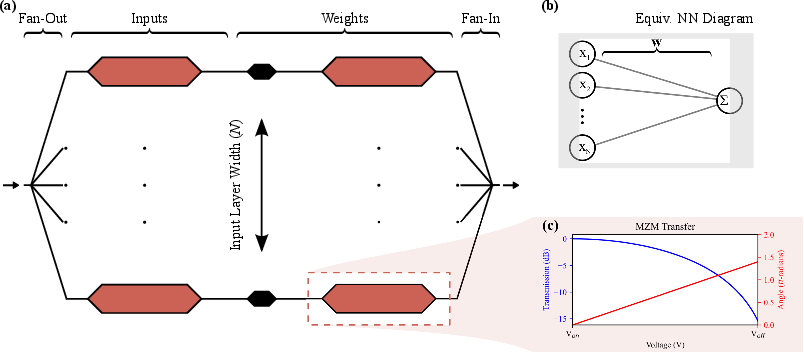}
    \caption{
    (a) The COLN architecture. The carrier signal is fanned out, where it meets an MZM that imparts the input and then meets another MZM that impart the magnitude of the weight, with the phase shifter controlling the sign of the weight. Each bus is then recombined at the Fan-In stage, representing the vector-vector multiplication of the COLN.
    (b) An equivalent NN diagram for this circuit. The unshaded region represents the multiplication and summation performed by the COLN, ignoring the activation function which is external to this architecture.
    (c) A typical transfer function for a MZM. The extinction ratio (ER) and phase can vary with design. 
    }
    \label{fig: trad COLN}
\end{figure}

We begin our design derivation with inspiration from the COLN \cite{mourgias2019neuromorphic}. The operating principle of the COLN employs four key stages: Fan-Out, Input, Weighting, and Fan-In, as shown in Figure \ref{fig: trad COLN}(a). At the Fan-Out, a continuous-wave (CW) optical carrier signal is split onto $N$ copies. Here, the Input and Weight stage is represented by amplitude modulators, which are implemented in hardware as Mach-Zehnder Modulators (MZM) and impose the inputs and weights onto the CW carrier signal. This signal is recombined from $N$ paths down to 1 path, performing summation of the $N$ input/weight products at the Fan-In through constructive interference. Figure \ref{fig: trad COLN}(b) demonstrates the equivalent neural network (NN) diagram for this circuit, which performs the linear summation of the inputs multiplied by the synaptic weights, preparing the carrier signal for the activation stage.
This optical circuit and the NN both reduce to the general form:
\begin{equation}\label{eq: MAC MZM}
    out = \sum_n^{N}\mathbf{w}_n\mathbf{x}_n, 
\end{equation}
where, for the optical circuit, $out$ is the electric field output, $N$ as the number of layers in the width (or Fan-Out), $\mathbf{w}$ is the weight vector, and $\mathbf{x}$ is the input vector. In this formulation, $\mathbf{w}$ and $\mathbf{x}$ are represented by the electro-optic transfer function of the MZM in each layer similar to that seen in Figure \ref{fig: trad COLN}(c). A key feature of the COLN circuit is that the output is in a state which, with minimal additional passive circuitry, is fully compatible both with an activation based on optical-electrical (OE), electrical processing, and electrical-optical (EO) conversion \cite{williamson2019reprogrammable}, direct optical-electrical-optical (OEO) activation \cite{tait2019silicon} or all-optical, non-linear activation \cite{jha2020reconfigurable, mourgias2019all}.

In a fiber-optic based system, phase-shifters and splitters are readily available off-the-shelf \cite{mourgias2019neuromorphic}. However, when attempting to scale these down for an integrated photonic system, the size of MZM translates into a practical barrier. Current PN diode, free-carrier MZM devices each occupy a physical space of $> 1$ mm$^2$ in on-chip area, severely limiting the scalabilty of this architecture. 
The first on-chip demonstration of the COLN in a silicon photonic chip utilizes a $25$ mm$^2$ chip to implement a $4$ input/weight linear neuron, implementing the weights with thermal phase shifters, which are dramatically smaller at the cost of update speed \cite{giamougiannis2021silicon, mourgias2020neuromorphic}.
\begin{figure}
    \centering
    \includegraphics[width = \textwidth]{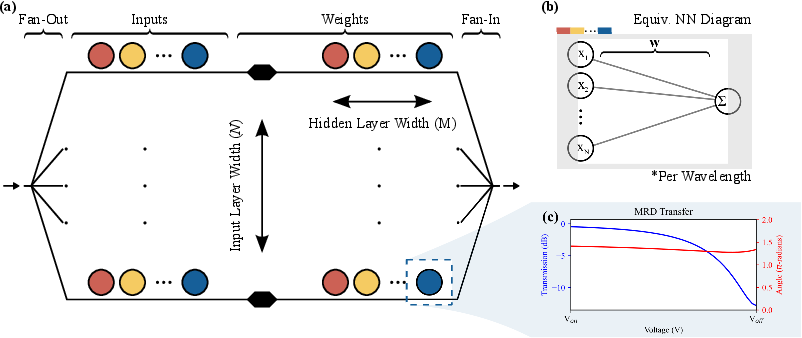}
    \caption{
    (a) The na\"ive  WDIPLN architecture. The carrier signal is fanned out, where it meets an MRD that imparts the input and then meets another MRD that imparts the magnitude of the weight, with the phase shifter controlling the sign of the weight. Each bus is then recombined at the Fan-In stage, representing the vector-vector multiplication of the WDIPLN. 
    (b) An equivalent NN diagram for this circuit. The unshaded region represents the multiplication and summation performed by the WDIPLN, ignoring the activation function which is external to this architecture. The tabs in the upper left hand side represent the WDIPLN's ability to represent $M$ many-to-one networks shown here. 
    (c) A typical transfer function for a MRD. The ER and phase can vary with design, however, in the ``slightly undercoupled'' regime we will see a relatively flat phase response. }
    \label{fig: trad WDIPLN}
\end{figure}

\begin{figure}
    \centering
    \includegraphics[width = \textwidth]{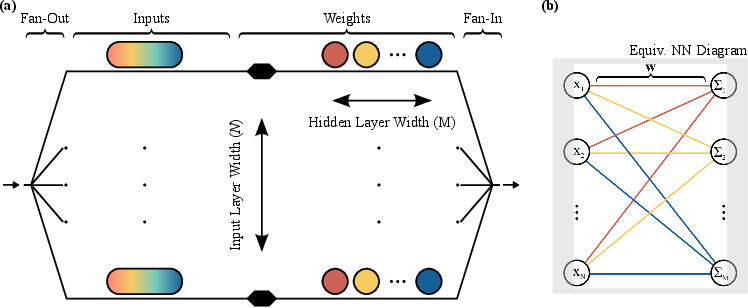}
    \caption{
    (a) The fully formed WDIPLN architecture. The carrier signal is fanned out, where it meets a large MRD that imparts the input to each channel simultaneously. Each channels input then meets a small MRD which imposes the magnitude of the weight, with the phase shifter controlling the sign of the weight. Each bus is then recombined at the Fan-In stage, representing the vector-matrix multiplication of the WDIPLN. 
    (b) An equivalent NN diagram for this circuit. The unshaded region represents the multiplication and summation performed by the WDIPLN, ignoring the activation function which is external to this architecture. This architecture fully enables the linear stage of MLP layer. }
    \label{fig: opt.WDIPLN}
\end{figure}

Our circuit architecture, which we subsequently refer to as Wavelength Diverse Integrated Photonic Linear Neuron (WDIPLN), begins as a COLN  with na\"ive replacement of MZMs with micro-resonant devices (MRD), either micro-ring resonators or a micro-disks, as shown in Figure \ref{fig: trad WDIPLN}(a). 
With MRDs, we immediately benefit from the decrease in footprint. Single-channel MRDs are commonly $< 100 \ \mu $m$^2$ in size \cite{timurdogan2014ultralow}. In addition, careful design of the MRD allows for a desirable adaptation of the previous electro-optic transfer function, displayed in Figure \ref{fig: trad WDIPLN}(c). The MRD acts as an amplitude filter both in the input and weight stage of the network, for each bus in the layer width. MRDs are sensitive, but if the coupling condition is designed to be slightly under-coupled, it is possible to mitigate some of the phase sensitivity (for example, a depletion biased MRD would remain under-coupled during operation). 
This enables the coherent summation at the Fan-In stage to remain relatively stable for each channel. 

Na\"ive WDIPLN introduces a feature beyond the COLN, namely wavelength diversity. For sufficiently small radii MRD (or, conversely, sufficiently large free-spectral-range (FSR)), we can place additional MRDs at new resonant wavelengths (or along specific channels, similar to wavelength-division-multiplexing (WDM) in optical communications).
\textcolor{black}{For example, in silicon photonics a suffiently large channel spacing has been demonstrated with a $2.5 \ \mu$m radius microdisk modulator, with $> 55 \ \mu$m, full width at half maximum of $\sim 0.2$ nm, and high quality factor ($>8,000$), which indicates high channel isolation \cite{timurdogan2014ultralow}.}

Recently, a similar WDM scheme was proposed using MZMs, but the WDIPLN implementation with ring resonators have the advantage that the wavelength multiplexing/de-multiplexing is done intrinsically using the single device in a compact footprint \cite{totovic2022programmable}. 
We show a schematic for the na\"ive WDIPLN in Figure \ref{fig: trad WDIPLN}(a). This enables wavelength diverse operation of the coherent circuit \textit{at each wavelength channel,} such that we have an incoherent circuit which acts like a coherent operator at each channel ($\lambda$), with a slight adaptation of Equation \ref{eq: MAC MZM} to 
\begin{equation}\label{eq: MAC MRD}
    out_{\lambda} = \sum_n^{N}
    \mathbf{w}_{\lambda, n} \mathbf{x}_{\lambda, n} ,
\end{equation}
\textcolor{black}{We can claim Equation \ref{eq: MAC MRD} for WDIPLN design with sufficiently isolated channels, where the interaction of each resonance occurs only in and around that wavelength. As discussed above, this is achieved by using sufficiently small MRDs with large FSRs. Using isolated wavelength channels enables the resonant based circuit to act as any $m$ individual COLN circuit for $m \in M,$ where $M$ here represents the number of wavelength channels in operation \textit{and} the number of COLN circuits represented by a single WDIPLN circuit. Therefore,} 
this design not only leverages smaller footprint devices, but enables further footprint reduction of \textcolor{black}{a given system of linear neurons} from the limit of one optical circuit per preceptron to one optical circuit per fully connected multi-layer perceptron (\textit{MLP}) by using the input/weight pairs at each wavelength as shown in  Figure \ref{fig: trad WDIPLN}(b). 

In the na\"ive case, we have simply swapped each MZM with an MRD, such that the circuit requires $2NM$ MRDs, for layer width $N$ and wavelength channels (or layer depth) $M$. This case is the same as using $M$ COLN circuits, which is equivalent only in abstracted device count complexity. Thus, an actual network with COLNs would require $M$ copies of the circuit resulting in a larger implementation.  
\begin{table}[ht!]
\centering
\footnotesize
\begin{tabular}{@{}lcccc}
\br
\textbf{Architecture}&COLN&COLN&WDIPLN&WDIPLN\\
(variation)&(nominal)&(w/ thermal MZI)&(na\"ive)&(nominal)\\
\mr
\textbf{Input Element} &MZM&MZM&Small MRD&Large MRD \\
Approx. Element Size (mm$^2$) &$0.8$&$0.8$&$10^{-4}$&$10^{-2}$ \\
\mr
\textbf{Weight Element} &MZM&Thermal MZI&Small MRD&Small MRD \\
Approx. Element Size (mm$^2$) &$0.8$&$10^{-1}$&$10^{-4}$&$10^{-4}$ \\
\mr
\textbf{Scaling Rule} &$2NM$&$2NM$&$2NM$&$N(M+1)$ \\
\mr
\textbf{Physical Size (mm$^2$)} &\multicolumn{4}{c}{Component Size Only (no routing considered)} \\
$N=8, M=1$ & $12.8$ & $7.2$ & $1.6\times10^{-3}$ & $8.08\times10^{-2}$  \\
$N=8, M=8$ & $102.4$ & $57.6$ & $1.28\times10^{-2}$ & $8.64\times10^{-2}$  \\
\mr
\textbf{Electrical I/O} &\multicolumn{4}{c}{Consider 4 Electrical I/O per element (2 EO, 2 Thermal)} \\
$N=8, M=1$ & $64$ & $64$ & $64$ & $64$  \\
$N=8, M=8$ & $512$ & $512$ & $512$ & $288$  \\
\mr
\textbf{\textcolor{black}{Power Consumption} (mW)} &\multicolumn{4}{c}{Same Configuration (2 EO, 2 Thermal) } \\
%
%
%
%
\ \ \textbf{EO Power} - ($P_{On/Off}$)\\
\ \ \ \ $N=8, M=1$ & $2.72$   & $1.36$  & $1.6$  & $0.48$  \\
\ \ \ \ $N=8, M=8$ & $21.76$  & $10.88$ & $12.8$ & $3.84$  \\
\ \ \textbf{Thermal Power} - ($P_{\pi}$)\\
\ \ \ \ $N=8, M=1$ & $89.6$    & $89.6$       & $89.6$     & $25.2$  \\
\ \ \ \ $N=8, M=8$ & $716.8$  & $716.8$     & $716.8$   & $201.6$  \\
\br
\end{tabular}\\
\caption{\label{table}Table comparing the primary differences between the COLN and WDIPLN architectures. We consider the nominal COLN, a COLN where weights are done with thermal MZIs, the na\"ive WDIPLN and nominal WDIPLN. For physical size calculations, we only consider the footprint of individual elements and recognize additional routing, both optical and electrical, will be necessary for each design. The physical size \textcolor{black}{and power consumption values} were estimated from available foundry PDKs in \cite{AIMPDK, globalPDK, cornerstonePDK}. }
\end{table}
However, we recognize that the input stage MRDs are redundant. Therefore, to realize full WDIPLN we replace all $M$ input MRDs for a given layer with a carefully designed, larger radius MRD with an FSR that matches the channel spacing, as outlined in Figure \ref{fig: opt.WDIPLN}(a). This adjustment lowers the total device count to $N(M+1),$ which represents a dramatic improvement in device count scalability in addition to the physical size. As seen in Figure \ref{fig: opt.WDIPLN}(b), the WDIPLN circuit in this form represents the linear weighting and summation stage of a fully connected \textit{MLP}. 
\textcolor{black}{A caveat of the fully-realized WDIPLN architecture is the inability to realize phase shifting per weight. In this way, we must limit the weight attribution to $[-1, 0]$ or $[0, 1]$. The phase shifters in this architecture are here as a global change to all weights simultaneously, serving both as path balancing and switching between subtractive or additive behavior.}
Table \ref{table} summarizes the key differences between the COLN and WDIPLN, including scaling rules, physical size, electrical I/O, and power consumption estimates from available foundry process design kits (PDKs) \cite{AIMPDK, globalPDK, cornerstonePDK}. We note that the physical size is calculated purely as fundamental device area, although both the COLN and WDIPLN require additional optical and electrical routing to connect to the outside world. This routing footprint was omitted from the total since it is highly implementation-specific and depends heavily on packaging constraints (i.e., edge coupled versus grating coupled, flip-chip bonded or wire-bonded, etc.). However, is is clear that the extremely small device footprint of the WDIPLN architectures leave ample room for additional circuitry, such that optical couplers and electrical bond pads will dominate the final device area. For example, $288$ electrical pads ($60~\mu$m $\times$ $60~\mu$m) on a $16 \times 18$ grid at $150~\mu$m spacing take up 6.48 mm$^{2}$ in on-chip area. If we add optical couplers ($400~\mu$m $\times$ $20~\mu$m) on the edge of the chip at a large pitch for ease of packaging, the total on-chip area is $<7$ mm$^{2}$ for the WDIPLN of $N=8, M=8$. 
\textcolor{black}{In addition, we note the power consumption values are also simply estimated from available information in \cite{globalPDK}, which utilize thermal isolation methods to achieve a $P_{\pi} \sim 2.8$ mW. Importantly, these values serve to illustrate the improved power budget of the WDIPLN archiecture -- regardless of technology implementation.}

\section{Experimental Demonstration of Simple Addition}
\begin{figure}
    \centering
    \includegraphics[width = \textwidth]{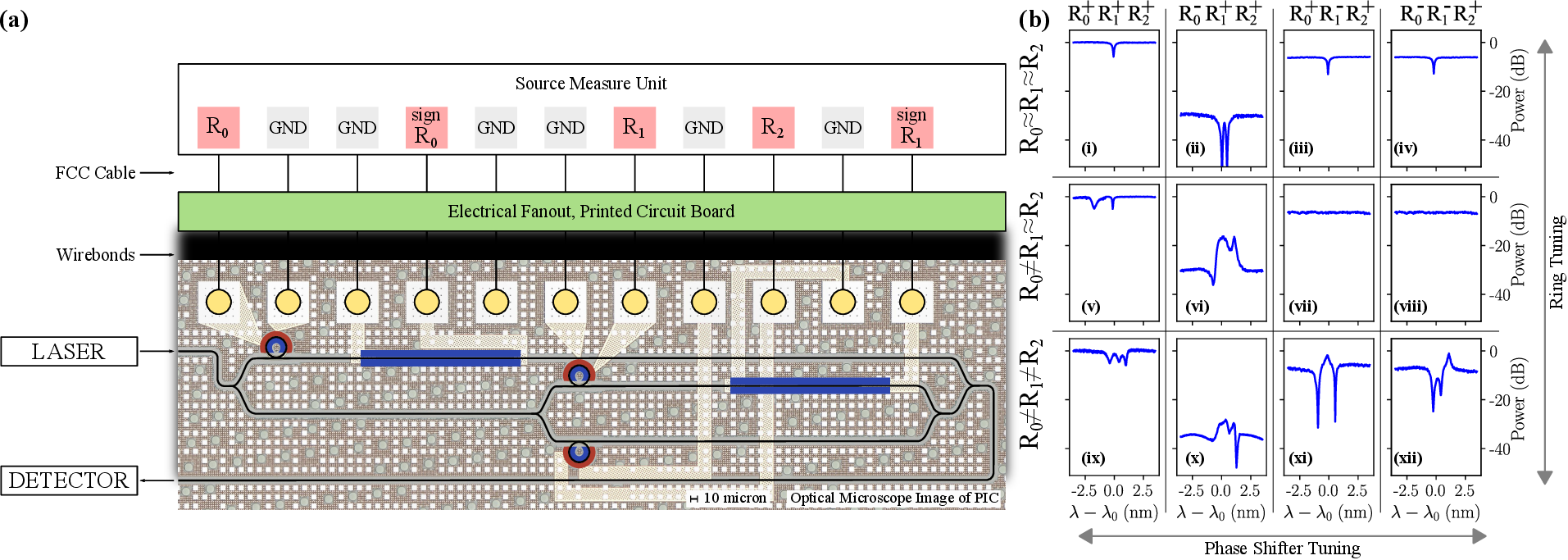}
    \caption{(a) Experimental setup. We optically couple the laser and detector to the photonic integrated circuit (PIC) using SMF28 fiber. The PIC consists of a simplified WDIPLN circuit with a bias line, for a total of 3 MRDs ($R_0, R_1, R_2$) and two phase shifters ($\textrm{sign}(R_0), \textrm{sign}(R_1)$. \textcolor{black}{All of the splitters/combiners are designed for equal splitting ratio.} The electrical traces connect the device terminals to the bond pads, which are wirebonded from the PIC to an electrical fanout and then connected to a printed circuit board. The printed circuit board is routed to an source measure unit (SMU) through a flat conductor cable (FCC). The SMU enables electro-optic control of each device. 
    (b) Demonstration of addition and subtraction. The columns here represent the tuning of the phase shifters in (a), and the rows represent the tuning of the MRDs. 
    The four columns show the following behavior: $+R_0 + R_1 + R_2$, $-R_0 + R_1 + R_2$, $+R_0 - R_1 + R_2$, $-R_0 - R_1 + R_2$. 
    The three rows demonstrate states where $R_0 \sim R_1 \sim R_2$, $R_0 \neq R_1 \sim R_2$, $R_0 \neq R_1 \neq R_2$.
    \textcolor{black}{The small wavelength range is kept to provide a ``big-picture'' of the circuit behaviour, however, we evaluate each operation at $\lambda = \lambda_0$.}}
    \label{fig: addsub}
\end{figure}

As an initial demonstration, we show the ability of WDIPLN to add and subtract in a simplified form. This represents a fundamental building block of the architecture. While tuning the MRDs ultimately imposes the input/weight onto the circuit, each of the $N$ buses in the circuit are additionally reconfigurable via a phase shifter. The phase shifter rotates the phase of a given bus so that we are able to express the sign coefficient of the weights as either $-1$ or $1$. 
Bias plays a role in many NN architectures, effectively shifting the set point of the linear fit. Similarly, in this architecture, we can introduce a bias to change our set point. From an optical perspective, this allows us to validate the operation of the circuit \textcolor{black}{via direct measurement through a photodetector at the output \cite{mourgias2019neuromorphic}}. The summation operator occurs as optical interference, which will ``add'' (constructively interfere) when the phase is the same, and ``subtract'' (destructively interfere) when the phase is opposite. For an initial demonstration, we consider the optical experiment in Figure \ref{fig: addsub}(a). 
We designed an WDIPLN circuit with an added bias. There are five active circuit elements: the bias ring, $R_0$, bias phase shifter, $PS_0$, top arm ring $R_1$, top arm phase shifter $PS_1$ and bottom arm ring $R_2$. Each element is electrically connected via a wirebond to an electrical fanout, which is in turn wirebonded to a printed circuit board (PCB) and connected to a source measure unit (SMU). The circuit is optically coupled from two single mode fibers (SMF28) embedded in a titled fiber array to two grating couplers on either side of the circuit. A laser and detector pair are connected to the other ends of the respective SMF28 fibers. The MRD we employ is a carrier injection-based PIN micro-ring resonator. We employ a PIN for proof of principle due to the large wavelength change, however, future systems will utilize application specific, optimized MRDs operating in carrier depletion mode. The photonic integrated circuit (PIC) was fabricated in American Institute for Manufacturing (AIM) Photonics' 300 mm silicon photonics process \cite{AIMPDK}.

\textcolor{black}{We demonstrate a variety of circuit configurations as summarized in Figure \ref{fig: addsub}(b). We achieve these configurations by tuning the phase shifters to $0$ or $\pi$ (columns) and setting the rings at various resonant wavelengths (rows), for which we indicate the state of each ring by the label of $R_0, R_1,$ or $R_2$. The bias path is slightly shorter than the internal WDILPN, which means we can utilize the phase shifter and the natural wavelength-dependent interference to extract the set-point behavior. Sub-figures (\textit{i, v, ix}) show three states of addition with phase shifters tuned to $0$ where the rings are all aligned to a single resonance ($R_0 = R_1 = R_2$, \textit{i}), $R_1$ is tuned off ($R_1 \neq R_0 = R_2$, \textit{v}), and all the rings are at different resonances ($R_0 \neq R_1 \neq R_2$, \textit{ix}). These states manifest as one, two, or three resonance peaks, respectively. Sub-figures
(\textit{ii, vi, x}) demonstrate the corresponding states where the bias phase is at $\pi$ rather than $0$. Due to manufacturing variations, we observe behavior that deviates from the designed point. In (\textit{ii}), we expect no signal to pass through; however, due to path and y-branch imbalance, the circuit floor is $\sim -30$ dB—additionally, small differences in the ring are seen in the transmission. Therefore, the maximum value in (\textit{ii}) is $\sim -33$ dB at $\lambda_0$, where the resonant wavelength sits, which indicates a slight imbalance but overall high suppression of any difference in the rings. Sub-figure (\textit{vi}) shows two peaks above the minimum, as $R_0 \neq R_1 = R_2$, and (\textit{x}) shows three distinguishable peaks as $R_0 \neq R_1 \neq R_2$. The peaks are overlapping such that distinguishing individual peaks is difficult. Sub-figures
(\textit{iii, iv}) demonstrate a single resonance, similar to (\textit{i}); however, note the maximum power is roughly $-6$ dB \textcolor{black}{since no light passes through in the lower path and each y-branch contributes additional loss of $3$ dB}. Sub-figures
(\textit{vii, viii}) show detuned $R_0$, and $R_1 = R_2$, where these rings fully cancel out each others' magnitudes (subtraction). Sub-figures
(\textit{xi, xii}) demonstrate $R_0 \neq R_1 \neq R_2$ for the subtraction operator, which is seen by the distinct peak difference above and below the bias line. Finally, $R_1, R_2$ each flip signs between (\textit{xi, xii}) due to the phase control of the bias line.}

\section{Experimental Demonstration of Logic Gates}
\begin{figure}
    \centering
    \includegraphics[width = 0.95\textwidth]{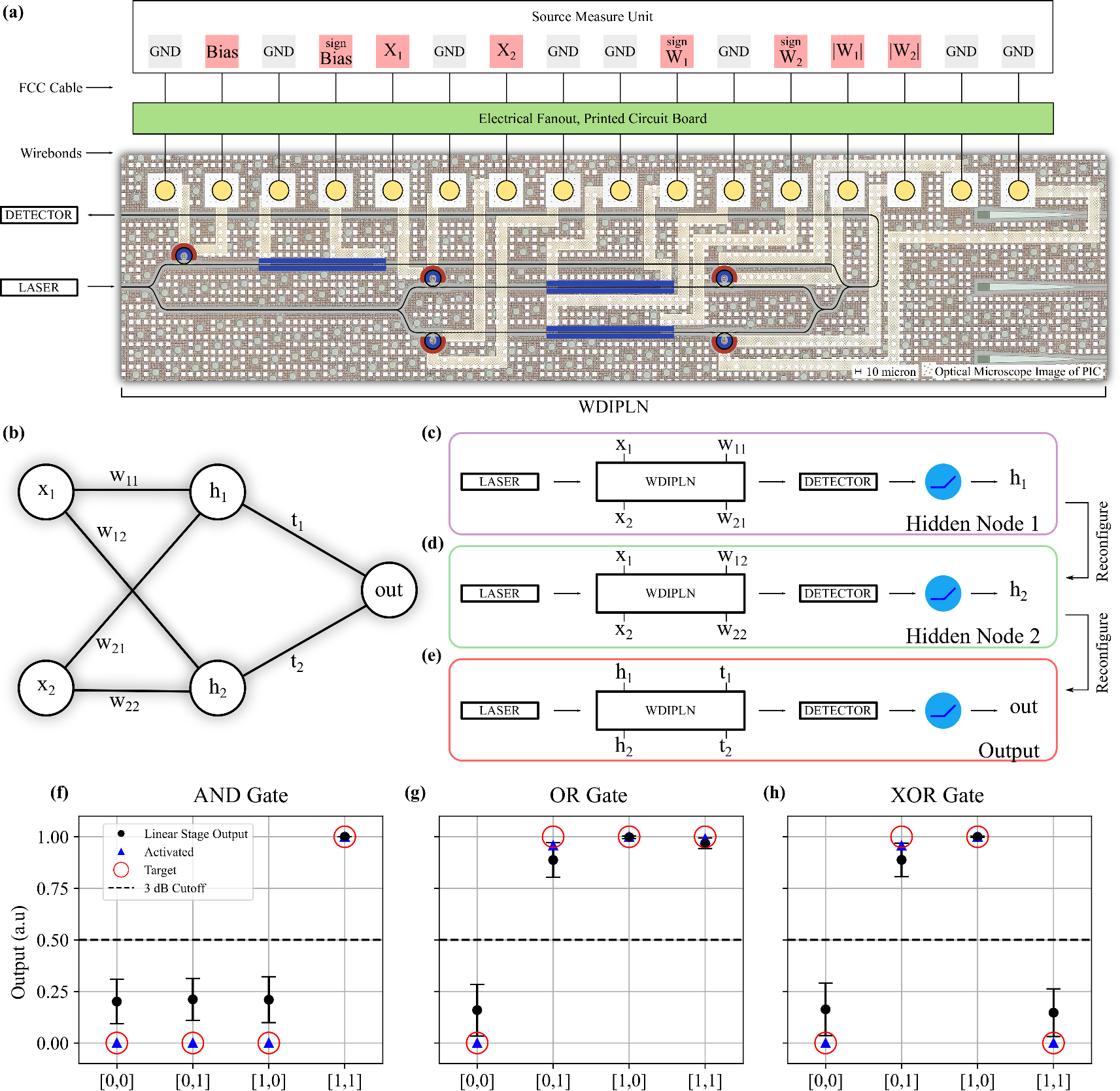}
    \caption{Photonic Neural Network experimental demonstration.
    (a) Experimental set-up, similar to Figure \ref{fig: addsub}. We optically couple the laser and detector to the PIC. The PIC consists of a WDIPLN circuit with $N = 2, M = 1$ and a bias line, for a total of 5 MRDs and 3 phase shifters. \textcolor{black}{All of the splitters/combiners are designed for equal splitting ratio.}The electrical wiring goes out the bond pads, which are wirebonded from the PIC to an electrical fanout and then connected to a printed circuit board. The printed circuit board is routed to a source measure unit (SMU) through a flat conductor cable (FCC). The SMU enables electro-optic control of each device. 
    (b) The selected architecture for the simple neural network we implement for this task. Inputs ($x_1, x_2$) are connected by a weight matrix, $\mathbf{w}$, to a hidden layer with nodes $h_1, h_2.$ From here, a simple weight vector, $\mathbf{t}$, connects the hidden layer to the output. 
    (c--e) The configure-recycle process for the circuit in (a). We send in the input pairs $[0,0], [0, 1], [1,0], [1,1]$ as $x_1, x_2$ for the first two stages, and subsequently send in $h_1, h_2$ in the final stage.
    We configure the weights of the WDIPLN to match that in (b) step-by-step from pairs [$w_{11}, w_{21}$], [$w_{12}, w_{22}$], and [$t_{1}, t_{2}$] for each stage of the network, respectively. Each configure-to-measurement cycle takes approximately 1 second, which is dominated by the read/write speeds of the SMUs. 
    (f--h) Results of the experimental demonstration. The AND, OR and XOR gate correctly predict the outputs with accuracy $96.8\%, 99\%,$ and $98.5\%$, respectively.
    }
    \label{fig: bitgate}
\end{figure}

We designed a second circuit in order to demonstrate a learning task. This design reflects the WDIPLN architecture for $N=2, M=1$, as seen in Figure \ref{fig: bitgate}(a). This circuit places both input and weight rings along the inner buses, along with a bias branch that also contains a ring. We wirebond out to an electrical carrier, which we then connect to programmable SMU channels in order to electrically control each element. We optically couple the laser and detector through fiber to grating couplers on the input and output waveguides, respectively.

We selected a simple network architecture that is able to reconfigure the weights in order to represent AND, OR and XOR gates, shown in Figure \ref{fig: bitgate}(b). We trained these networks on a CPU using PyTorch \cite{NEURIPS2019_9015}. For the activation function we employ the rectified linear-unit ($\textrm{ReLU}$) for all stages.
We can create a simple map from the trained model to our physical chip by reconfiguring the chip in between the three linear stages, leading to the three non-input nodes $h_1, h_2$ and $out$. For the ring resonators, we define a value of ``0" as $V_{off} = 1.4$ V and ``1'' as $V_{on} = 1.2$ V. This ensures we are operating in the slightly undercoupled regime for both ``0'' and ``1''. We note that for this demonstration, the $V_{off}$ and $V_{on}$ values are globally set for all rings. In addition, the bias ring is set to $V_{off}$ and the bias phase shifter is set to $\pi/2$, \textcolor{black}{specifically for visibility of the full range of $[-1,1]$ through direct detection as opposed to inference \cite{mourgias2019neuromorphic}}.

The process of reconfiguration and implementation of the two hidden nodes and the output is shown in Figure \ref{fig: bitgate}(c--e).
For the two hidden nodes, $h_1$ and $h_2$, we configure the weights of the optical circuit, namely $|W_1|$ and $|W_2|$, according to the pairs of $[w_{11}, w_{21}]$ and $[w_{12}, w_{22}]$, respectively. In addition, if the sign of the weight is negative, we adjust the corresponding phase shifter, namely $\textrm{sign}(W_1)$ and $\textrm{sign}(W_2),$ from $0$ to $\pi$. Once we configure the chip for $h_1$, we feed the four input pairs $[0,0], [0, 1], [1,0], [1,1]$ into $X_1, X_2$. For each input, we measure the output from the detector, apply the ($\textrm{ReLU}$) activation function used in training with a ``$3$ dB cutoff'', and record the result. After the two hidden nodes are measured, we reconfigure the circuit for the final stage, according to the final weight stage from training $t_1, t_2$. Finally, we feed the four recorded pairs of $h_1, h_2$ into the circuit as inputs, measure the output, apply the activation function and record the result. 

According to the training-to-implementation scheme described above and in Figure \ref{fig: bitgate}(c--e), we demonstrate the circuit's performance for three different 2-bit logic gates, namely AND, OR and XOR. 
In training, the accuracy of this simple network reaches $\sim 100\%$. The resulting outputs of the three gates are shown in Figure \ref{fig: bitgate}(f--h). The black dots are the square root (i.e. magnitude) of the output values measured by the detector at the laser probe wavelength, which in this experiment was globally set to $\lambda_0 = 1,526$ nm. The error bars represent the standard output deviation for a $100$ pm window around, $\lambda_0 \pm 50$ pm. The blue triangles show the outputs of the circuit after activation, and the red circles show the target for the gates. The AND, OR and XOR gates achieve predictive accuracy of $96.8\%, 99\%,$ and $98.5\%$, respectively.

\section{Conclusion}
Silicon photonics holds promise for implementing large-scale PNNs directly on chip with substantially improved performance compared to electronic implementations. In this work, we propose and demonstrate a novel PNN architecture, WDIPLN, which exhibits highly favorable characteristics compared to previous literature. The WDIPLN derives from the COLN architecture detailed in \cite{mourgias2019neuromorphic}, enabling similar vector-vector multiplication and summation using an optical carrier signal with the additional benefits of wavelength parallelism, drastically reduced device footprint, and significantly lower energy consumption. The majority of these beneficial properties come from exchanging large, power-hungry MZMs with compact MRDs. While we can trivially exchange the fundamental elements (MRDs for MZMs) in a one-to-one manner to enable multi-channel operation, we show that by replacing the input MRDs with a single, multi-channel MRD with an FSR equal to the weighting channel spacing we can reduce the total device count significantly. As a first proof-of-principle, we experimentally demonstrated simple addition and subtraction between rings in a WDIPLN circuit. We then show a WDIPLN configured with $N = 2$ and $M = 1$, which we configure and recycle to perform 2-bit logic gates (AND, OR and XOR). We observe the implementation accuracy for each gate to be $96.8\%, 99\%,$ and $98.5\%$, respectively. Furthermore, the natural wavelength parallelism of the proposed WDIPLN architecture can be exploited using chip-based Kerr frequency comb sources \cite{Gaeta2019} for massive scaling in the frequency domain, similar to recent demonstrations in the silicon photonics platform for high bandwidth data communications \cite{rizzo2021integrated}. This demonstration opens new opportunities in massively parallel silicon photonic PNNs and paves the way to large-scale systems in the thousand-neuron regime on a single chip with minimal energy consumption.

\section*{Data Availability}
The data that support the findings of this study are available upon request from the authors.

\ack

This work was supported by Air Force Research Laboratory (FA8750-21-2-0004 and FA8650-21-2-1000) and the National Science Foundation (Award \#1810282). The views and conclusions contained herein are those of the authors and should not be interpreted as necessarily representing the official policies or endorsements, either expressed or implied, of the United States Air force, the Air Force Research Laboratory or the U.S. Government. The U.S. Government is authorized to reproduce and distribute reprints for Governmental purposes notwithstanding any copyright notation thereon. M.v.N is funded by NSF award DMR-1747426. The authors thank AIM Photonics for chip fabrication. 

\appendix

\section{Operating Conditions and Variation of MRDs}

The mechanism of tuning for MRDs here is assumed to be electro-optic (where thermo-optic is an accepted simplification) and well defined by free-carrier plasma dispersion \cite{soref1987electrooptical}. The transfer function from the electrical to optical domain is defined as
\begin{equation}\label{eq: ring-transfer}
    E_{\textrm{all-pass-ring}} (\lambda; V) = e^{i (\pi + \theta(\lambda, V))} \frac{a(\lambda,V) - r e ^{-i\theta(\lambda, V)}}{1 - r a(\lambda,V) e^{i\theta(\lambda, V)}}, 
\end{equation}
where $a$ the voltage (V) dependent absorption (or loss) in the cavity lumped with intrinsic cavity loss, $\theta$ is the wavelength- and voltage-dependent phase shift, $r$ is the self-coupling coefficient (with a corresponding cross coupling coefficient, $\kappa$, such that $\kappa^2 + r^2 = 1$). 
Equation \ref{eq: ring-transfer} depicts the coupled effect of tuning, which both changes the loss and the relative cavity size (i.e. phase shift). 

\begin{figure}[ht!]
    \centering
    \includegraphics[width = \textwidth]{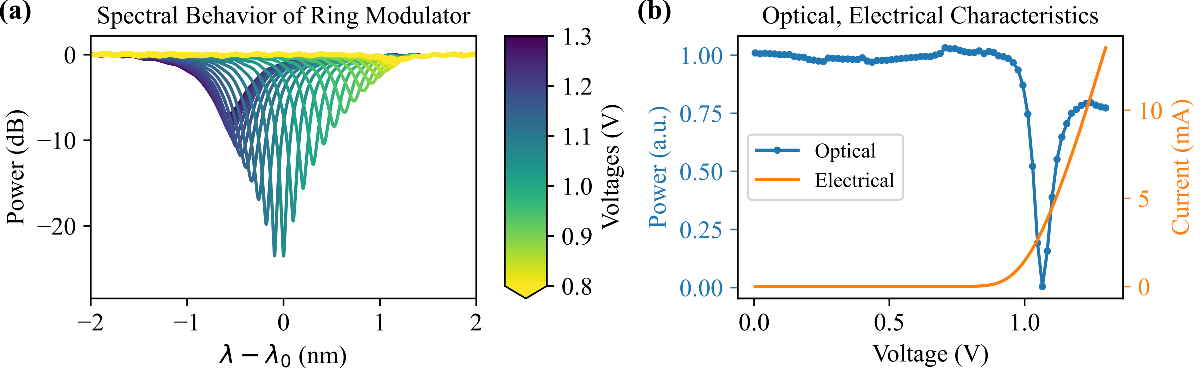}
    \caption{
    (a) The spectral behavior of the ring modulator used in our demonstration. As the voltage increases, the resonant wavelength blue-shifts to a lower wavelength. In addition, we see that the ring traverses the initial condition (over-coupled), to critical coupled, finally to under-coupled. This ring was selected due to the large tuning range for initial demonstration.
    (b) Optical and electrical characterization of the ring. A single wavelength, $\lambda - \lambda_0 = 0$ from (a), is plotted to show the change in optical transmission as a function of applied bias. From an electrical perspective, we see that the operating regime for this diode is indeed in forward-bias, with a maximum power for full tunability of approximately $12$ mW. 
    }
    \label{fig: app.ring_charac}
\end{figure}

In this work, we employ carrier injection-based ring resonators through the well-known free carrier plasma dispersion effect in silicon. These rings demonstrate a large spectral tuning range, as seen in Figure \ref{fig: app.ring_charac}(a). However, this shift comes at the cost of change of coupling conditions due to the strongly voltage-dependent round-trip loss. Changing the coupling conditions from over coupled to critically coupled to under coupled has a large impact on the phase shift associated the ring. It is precisely this phase that we aim to leverage in this architecture. Therefore, we operate the rings in the slightly-to-moderately under coupled regime in this experiment, which corresponds to voltages $V > 1.1~V$, as seen in Figure \ref{fig: app.ring_charac}(b). 

\begin{figure}[ht!]
    \centering
    \includegraphics[width = \textwidth]{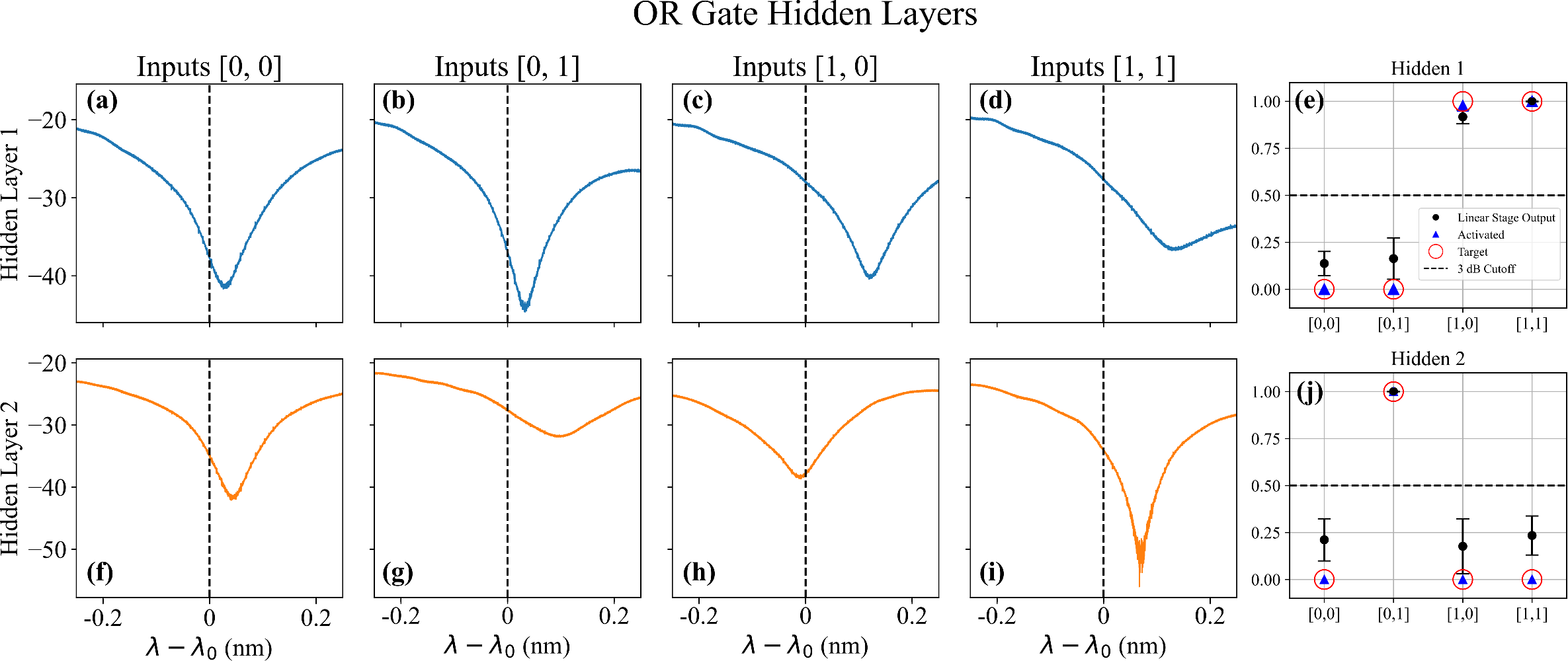}
    \caption{Full implementation sequence for the hidden layers of the OR gate. 
    (a--d) The circuit response when configured to the trained weights of hidden node 1 for the incoming inputs $[0,0], [0, 1], [1,0],$ and $[1,1]$.
    (e) Circuit output for Hidden Node 1 at the probe wavelength, with a range of $\pm 100$ pm.
    (a--d) Show the circuit response when configured to the trained weights of Hidden Node 2 for the incoming inputs $[0,0], [0, 1], [1,0],$  and $[1,1]$.
    (e) Circuit output for Hidden Node 2 at the probe wavelength, with a range of $\pm 100$ pm.}
    \label{fig: app.ORgatesample}
\end{figure}

Additionally, this demonstration did not employ thermal tuners to shift the rings coarsely. This leaves the demonstration susceptible to fabrication variations. For large-scale demonstrations, these thermal phase shifters can be implemented with isolated heaters with high efficiency \cite{coenen2022thermal, vanniekerk2022wafer} in concert with reverse-biased MRDs for low-power electro-optic configuration \cite{timurdogan2014ultralow, vanniekerk2022high}.

\section{Hidden Layer Demonstration Results}

The hidden layers are not explicitly necessary in order to run the network since the final stage weights are known. However, they are important in demonstrating the full generality of the WDIPLN architecture, including the ability to reconfigure. In Figure \ref{fig: app.ORgatesample}, we show an in-depth look at the procedure of validating the hidden nodes, $h_1$ and $h_2$. First, we configure the circuit to measure $h_1,$ as shown in Figure \ref{fig: bitgate}(c). Figure \ref{fig: app.ORgatesample}(a--d) show the optical circuit response in the Hidden 1 configuration for the four logic inputs, with (e) showing the outputs at the probe wavelength. Figure \ref{fig: app.ORgatesample}(f--i) shows the optical circuit response in the Hidden 2 configuration for the four logic inputs, with (j) showing the outputs at the probe wavelength.





\section*{References}
\bibliographystyle{iopart-num}
\bibliography{bib}

\end{document}